\documentclass{article}

\usepackage{siunitx}
\usepackage{PRIMEarxiv}
\usepackage{svg}
\usepackage[utf8]{inputenc} 
\usepackage[T1]{fontenc}    
\usepackage{hyperref}       
\usepackage{url}            
\usepackage{booktabs}       
\usepackage{amsfonts}       
\usepackage{nicefrac}       
\usepackage{microtype}      
\usepackage{lipsum}
\usepackage{fancyhdr}       
\usepackage{graphicx}       
\graphicspath{{media/}}     

\title{Semantically consistent Video-to-Audio Generation using Multimodal Language Large Model
}

\author{
Gehui Chen$\phantom{}^{1}$\thanks{Equal Contribution.}\hspace{10pt}
Guan'an Wang$\phantom{}^{\dagger}$\thanks{Project Lead.}\hspace{10pt}
Xiaowen Huang$\phantom{}^{1,2,3}$\thanks{Corresponding Author.}\hspace{10pt}
Jitao Sang$\phantom{}^{1,2,3}$
\and
$\phantom{}^1$School of Computer Science and Technology, Beijing Jiaotong University
\and
$\phantom{}^2$Beijing Key Lab of Traffic Data Analysis and Mining, Beijing Jiaotong University
\and
$\phantom{}^3$Key Laboratory of Big Data \& Artificial Intelligence \\
in Transportation(Beijing Jiaotong University), Ministry of Education
}

\makeatletter
\def\@fnsymbol#1{\ensuremath{\ifcase#1\or \dagger\or \ddagger\or *\or
   \mathsection\or \mathparagraph\or \|\or **\or \dagger\dagger
   \or \ddagger\ddagger \else\@ctrerr\fi}}
    \makeatother
\begin{document}
\maketitle{}

\begin{abstract}
Existing works have made strides in video generation, but the lack of sound effects (SFX) and background music (BGM) hinders a complete and immersive viewer experience. We introduce a novel \underline{s}emantically consistent \underline{v}ideo-to-\underline{a}udio generation framework, namely \textbf{SVA}, which automatically generates audio semantically consistent with the given video content. The framework harnesses the power of multimodal large language model (MLLM) to understand video semantics from a key frame and generate creative audio schemes, which are then utilized as prompts for text-to-audio models, resulting in video-to-audio generation with natural language as an interface. We show the satisfactory performance of SVA through case study and discuss the limitations along with the future research direction. The project page is available at \url{https://huiz-a.github.io/audio4video.github.io/}.
\end{abstract}

\keywords{video-to-audio generation \and multimodal language model \and text-to-audio generation}

\section{Introduction}
Recent video generation works like Sora\cite{videoworldsimulators2024} have exemplified notable breakthroughs within this domain. However, these generated videos are mute, lacking SFX and BGM, may cause a less enjoyable experience for the audience. In comparison with manual sound design and music scoring, automatic generation methods for audio and music harmonizing semantically with the video content presents a more efficacious and adaptable methodology, consequently refining the pipeline of video generation.

Currently, Significant advancements have been achieved in audio and music generation. For instance, AudioGen\cite{kreuk2022audiogen} and MusicGen\cite{copet2023simple} exhibit the capability to process textual inputs and synthesize audio or music that semantically align with the provided text. Pika\cite{pika2024} employs text-based sound generation techniques, enabling users to design prompts tailored to their videos for desired SFX. Nevertheless, this approach remains semi-automated as it necessitates prompt engineering. It is obvious that automatic prompt generation matching video semantics could boost audio generation efficiency. 

Specifically, we can harness the robust image understanding capabilities of MLLM\cite{wu2023multimodal} to comprehend the semantic content in videos through the analysis of key frames within video streams. Moreover, with extensive world knowledge and reasoning abilities\cite{wu2023multimodal}, MLLM is then capable of generating creative and content-consistent SFX and BGM schemes, which serve as prompts for audio generation models. Accordingly, we develop an efficient framework named SVA (semantically consistent video-to-audio generation), utilizing natural language as an interface for the automatic audio generation based on video content. Noted that audio generation may encounter challenges such as subpar quality and noise interference, thereby requiring post-processing methods such as denoising and mixing to attain the final results.

In this work, we present an efficient and straightforward framework aimed at achieving automatic audio generation based on video semantics. The framework comprises the following key steps:
\begin{itemize}
    \item Utilizing MLLM for comprehending video content and generating audio and music schemes.
    \item Employing generation models to produce audio or music in line with the given schemes.
    \item Incorporating fast noise detection, noise reduction, and mixing methods to generate high-quality videos with audio.
\end{itemize}
In this paper, We will provide our solutions to these key steps and discuss the limitations and future research direction.

\section{Method}
\label{sec:headings}
In the video-to-audio generation, the user will provide a silent video, and the model needs to produce audio that aligns with the semantic content of the input video. 

\begin{figure}[tb]
  \centering
  \includegraphics[width=0.8\textwidth]{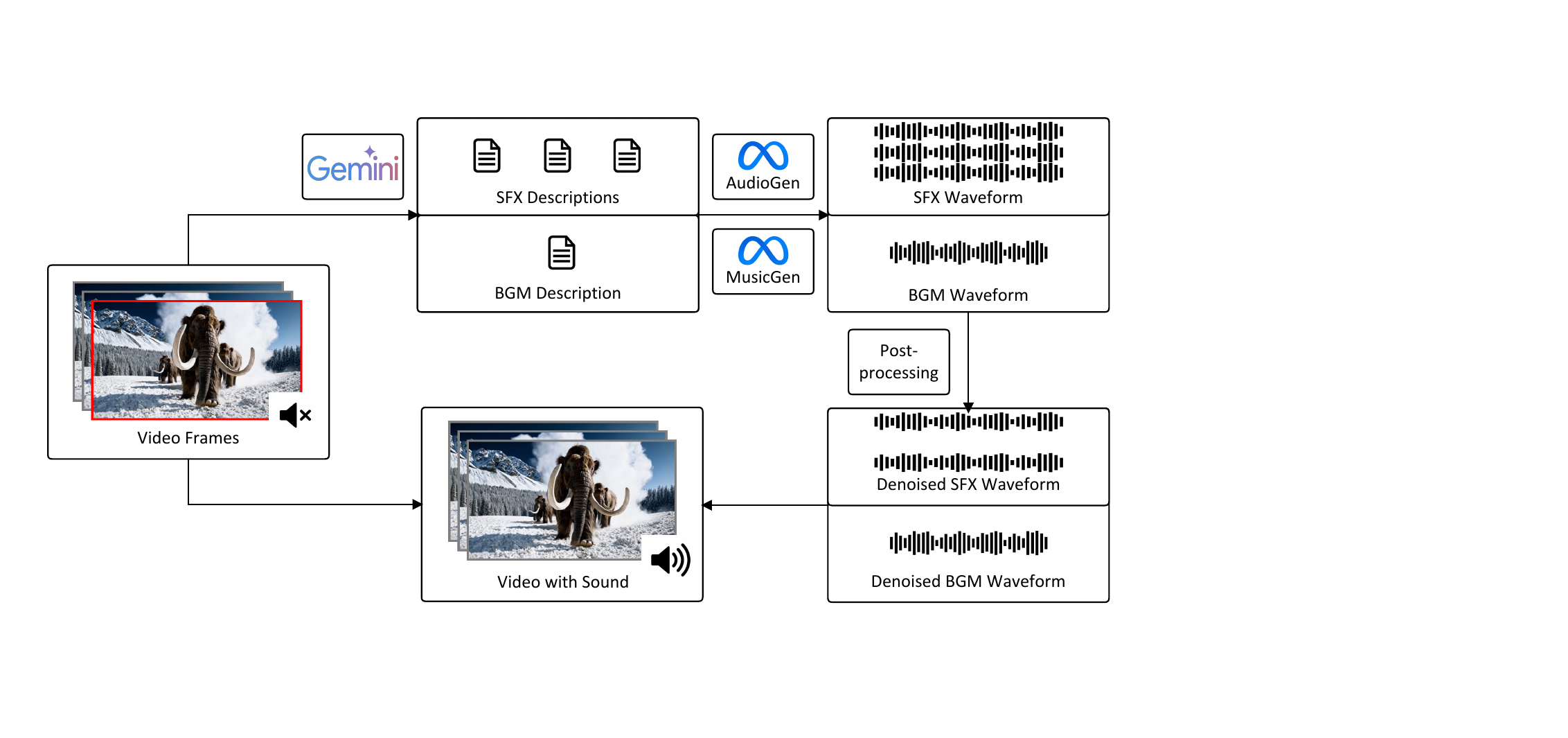}
  \caption{The overview of SVA framework. Initially, a key frame (highlighted in red) is randomly selected from the video frames. We then prompt Gemini Pro to generate a SFX and BGM scheme, which comprising two different SFX descriptions and one BGM description. Subsequently, these descriptions are inputted into AudioGen and MusicGen respectively, resulting in corresponding SFX and BGM waveform files. Then we run post-processing for noise removal and reduction. Finally, all multimodal data, including video frames, SFX waveform files, and one BGM waveform file are merged to create the video with audio.}
  \label{fig:fig1}
\end{figure}

Direct video-to-audio generation presents a considerable challenge. Given the advanced state of language-based multimodal technologies, we employ an indirect approach, utilizing text as a mediator for conveying semantic information and establishing a connection between the video and audio modalities. Therefore we propose a framework, SVG, which features a collaboration between MLLM and text-guided audio generation models. In this partnership, MLLM acts as the semantic conveyor, tasked with comprehending the video content and generating semantically consistent  SFX and BGM descriptions as a scheme. Simultaneously, the audio generation models receive the scheme and generate corresponding SFX and BGM. The overview of the SVG is depicted in Figure \ref{fig:fig1}. 

The process of the audio generation can be formulated as follows: $v \rightarrow p \rightarrow a$, where $v$ represents the video frames as input, $p$ denotes the scheme generated by MLLM, which serves as the prompt for audio generation. $a$ signifies the generated audio. We separately handle this process in two steps: $v \rightarrow p $ and $ p \rightarrow a$, with text $p$ as a bridge. In this section, we will provide an in-depth exposition of the proposed SVA. 

\subsection{SFX \& BGM Scheme}
Configuring appropriate SFX and BGM necessitates a profound comprehension of the video content and a semantically consistent scheme as prompts for the text-guided audio generation. MLLM possesses the capability to address both tasks: video semantic understanding and scheme generation. 

\textbf{Key Frame Selection}. Understanding the full frames sequence is time-consuming and unnecessary, especially in the short videos setting where scene transitions are infrequent, such as the example videos provided by Sora\cite{videoworldsimulators2024}. Utilizing one key frame to represent the entire video frames not only conveys the semantics but also optimizes the framework's efficiency. In detail, we use FFmpeg\footnote{\url{https://ffmpeg.org/}} to extract key frames of \textit{v}, then randomly select one frame as the image input. 

\textbf{Video Content Understanding}. To accurately understand the video content, we employ Gemini Pro\cite{geminiteam2023gemini}, a state-of-the-art MLLM, to interpret the content of image inputs. We prompt it to elicit an image description as the video content description. See Figure \ref{fig:fig2} for the template of video content description.

\textbf{Scheme Generation}. Upon acquiring the description of the video content, we provide it to Gemini Pro, leveraging its rich world knowledge and reasoning capabilities to obtain a semantically consistent and creative SFX and BGM scheme. One scheme comprises two SFX descriptions and one BGM description, considering that videos typically involve multiple entities emitting sounds simultaneously, whereas background music remains singular. We design a prompt template using a few-shot approach, as illustrated in Figure \ref{fig:fig2}. , which provides specific details. 
\begin{figure}[tb]
  \centering
  \fcolorbox{black}{gray!10}{\parbox{.9\linewidth}{
  \textbf{Template for video content understanding:}
  
  You are a video content understanding robot, and you are very good at understanding the content of a video through video images. Here is a key frame extracted from a short video. Please carefully understand this image to describe the video content based on the picture captured from it.

  Output: \textbf{\textcolor{red}{<Description>}}
  
  \textcolor{lightgray}{\rule[2pt]{\linewidth}{0.9pt}}
  \textbf{Template for scheme generation:}

  BGM stands for Background Music. It refers to the musical soundtrack or audio accompaniment that plays in the background of various media... SFX stands for Sound Effects.  It refers to all the sounds in a film, video game, or other media besides dialogue and music.  These sounds can be...
~\\

        You are good at coming up BGM and SFX ideas for a video based on its description. Here is the description of a video: \textbf{\textcolor{red}{<Description>}}. Now plan creative, wild and imaginative, blue-sky thinking and out of this world SFXs and BGM, in order to express innovation and amusement for the audiences.

        ~\\
        The SFX output is one short sentence in 12 words, which is not concerning music and instruments, and must be common in real world. If the sound is unusual and novel, find some similar and normal sound to replace it,like:

        \begin{itemize}
            \item 'Sirens and a humming engine approach and pass';
            \item 'A duck quacking as birds chirp and a pigeon cooing';
            \item ...
        \end{itemize}

        ~\\
        The BGM output is a brief description, involving instruments, beats, melody, mood, style and so on,like:
        \begin{itemize}
            \item 'A grand orchestral arrangement with thunderous percussion, epic brass fanfares, and soaring strings, creating a cinematic atmosphere fit for a heroic battle';
            \item 'Smooth jazz, with a saxophone solo, piano chords, and snare full drums';
            \item ...
        \end{itemize}

        ~\\
        Now output one SFX and BGM idea that creative, wild and imaginative, blue-sky thinking and Out of this world, comprising one BGM and 2 SFXs. Output the idea following the examples below in json, do not use any brackets and single quotes '' in a string: 
        \begin{itemize}
            \item \{"idea":"Mystical Curiosity",
        "SFX":["High-pitched wind chime tinkling softly","Distant owl hooting softly"],
        "BGM":"A whimsical and playful piece with a glockenspiel melody, light percussion using woodblocks and triangles, and a backdrop of ethereal chimes"\};
            \item \{"idea":"Prehistoric Dance Party"
        "SFX":"Stomping mammoth feet shaking the ground", "High-pitched trumpet calls from the mammoths"
        "BGM":"Upbeat electronic dance music with a strong bassline and prehistoric-inspired synth sounds"\};
            \item...
        \end{itemize}

        Output: \textbf{\textcolor{red}{<Scheme>}}
    }}        
  \caption{The template for prompting MLLM to generate video description and audio scheme. The process will be executed from top to bottom. The red-highlighted parts are placeholders, to be replaced according to the actual user input and MLLM output. Some non-critical content is omitted.}
  \label{fig:fig2}
\end{figure}

\textbf{User personalization}. The scheme generation template can be personalized to meet specific requirements if necessary. Users can freely describe their needs, such as "creating a melancholic atmosphere" or "an electronic music style solution." Then we instruct the MLLM to modify the template for scheme generation in Figure \ref{fig:fig2} according to the user's input, including few-shot examples and keywords replacement. For instance, if the user seeks a scheme imbued with a melancholic ambiance, the MLLM will substitute all few-shot examples with melancholic ones and embed terms like "melancholy" and "sadness" into the scheme generation template. Ultimately, the MLLM will furnish a scheme that evokes a somber emotional atmosphere while remaining congruent with the video content. The implementation details are shown in Figure \ref{fig:fig3}.
\begin{figure}[tb]
  \centering
  \fcolorbox{black}{gray!10}{\parbox{.9\linewidth}{
  \textbf{User Input:}
  
  \textbf{\textcolor{red}{<User Input>}}
  
  \textcolor{lightgray}{\rule[2pt]{\linewidth}{0.9pt}}
  \textbf{Template for substitutive few-shot examples:}
  
  Here are some SFX and BGM idea in json format. please follow the format and create more samples which satisfy the requirements: \textbf{\textcolor{red}{<User Input>}}

  \begin{itemize}
            \item {\{"idea":"Mystical Curiosity", "SFX":["High-pitched wind chime tinkling softly","Distant owl
hooting softly"], "BGM":"A whimsical and playful piece with a glockenspiel melody, light
percussion using woodblocks and triangles, and a backdrop of ethereal chimes"\}};
            \item {\{"idea":"Prehistoric Dance Party" "SFX":"Stomping mammoth feet shaking the ground", "High-
pitched trumpet calls from the mammoths" "BGM":"Upbeat electronic dance music with a strong
bassline and prehistoric-inspired synth sounds"}\};
            \item ...
    \end{itemize}

Now give the output satisfy the requirements: \textbf{\textcolor{red}{<User Input>}}. Just output without other content.

Output: \textbf{\textcolor{red}{<Examples>}}

\textcolor{lightgray}{\rule[2pt]{\linewidth}{0.9pt}}
\textbf{Template for substitive key word:}

A user is asking about a SFX \& BGM scheme for a short video.
Here is what the user said: \textbf{\textcolor{red}{<User Input>}}

 Now extract the key word indicating the user's requirement  like adj or noun. Don't mention SFX, BGM, short video because it's understood without needing to be said.

output example:
\begin{itemize}
            \item {Fresh, Upbeat, Cheerful, Lively, Sunny};
            \item {Electronic, Avant-garde, Technology, Experimental, Innovative};
            \item {Emotional, Memory, Cozy, Quiet, Melodic};
    \end{itemize}

Output: \textbf{\textcolor{red}{<Key Words>}}  

\textcolor{lightgray}{\rule[2pt]{\linewidth}{0.9pt}}
\textbf{Personalized template for scheme generation:}

...

~\\
        You are good at coming up BGM and SFX ideas for a video based on its description. Here is the description of a video: \textbf{\textcolor{red}{<Description>}}. Now plan \textbf{\textcolor{red}{<Key Words>}} SFXs and BGM, in order to express innovation and amusement for the audiences.
        ~\\
        ...
        
        ~\\
        Now output one SFX and BGM idea that \textbf{\textcolor{red}{<Key Words>}}, comprising one BGM and 2 SFXs. Output the idea following the examples below in json, do not use any brackets and single quotes '' in a string: 
        \begin{itemize}
            \item {\textbf{\textcolor{red}{<Examples>}}};
        \end{itemize}

        Output: \textbf{\textcolor{red}{<Scheme>}}
    }}  
  \caption{The template for personalized scheme generation. The process will be executed from top to bottom. The red-highlighted parts are placeholders, to be replaced according to the actual user input and MLLM output. Some non-critical content has been omitted.}
  \label{fig:fig3}
\end{figure}

\subsection{Text-to-Audio Generation}
MLLM accurately comprehends the video content and utilizes it to generate natural language descriptions of audio and music. Hence, the video-to-audio process can be perceived as a text conditional audio generation task. Being state-of-the-art open-source models in text-to-audio and music generation, AudioGen and MusicGen can produce high-quality audio semantically adhering to the provided text. To accomplish the video-text-audio process, we directly utilize the scheme provided by MLLM to prompt AudioGen and MusicGen. 

\subsection{Post-processing}
After obtaining SFX and BGM through the text-to-audio model, post-processing steps, including noise removal, reduction and mixing, are imperative to attain the ultimate high-fidelity video-to-audio output. 

\textbf{Noise Removal}. In practice, it has been noted that AudioGen can sometimes yield noise that entirely meaningless and devoid of useful information, whereas MusicGen consistently produces musically cohesive audio of superior quality. It is necessary to employ a method to detect and directly discard meaningless SFX audio files, as they are entirely unusable even after noise reduction. In order to effectively accomplish the aforementioned detection task, we run short-time Fourier (STFT) transform on the generated SFX. Then we compute the Root Mean Square (RMS), which is typically used to characterize the overall energy level of a signal\cite{10.5555/227373}. If the average energy exceeds a predefined threshold parameter, the audio file will be discarded. In this study, we employed the Librosa\footnote{\url{https://librosa.org/}} for conducting STFT and energy computation, with the threshold set at \num{0.3} based on preliminary experiments.

\textbf{Noise Reduction}. Following the removal of meaningless audio, we employ a combination of built-in noise reduction filters within FFmpeg to denoise the preserved audio. In particular, we use low-pass and high-pass filters to eliminate sounds below \num{200} Hz or above \num{3000} Hz, and configure the suppression level of the adaptive fast Fourier transform audio denoiser to \num{-25} decibels, resulting in a substantial reduction of noise in the SFX and BGM. 

\textbf{Video-Audio Merging}. Finally, we use FFmpeg's audio mixer filter to merge the denoised SFX and BGM into a single audio stream. This audio stream is then combined with the original video frames to produce the final video file.  Specifically, We apply the volume filter to adjust the SFX volume to \num{5}\% of its original level and the BGM volume to \num{3} times its original. Following this adjustment, we employ the amix mixer to amalgamate multiple audio streams into a unified audio stream. Original audio files are formatted as WAV. To integrate the audio into an MP\num{4} container, we employed an AAC encoder with a bitrate of \num{192}K, resulting in an audio-enabled video.

\section{Case Study}
\label{sec:others}
\begin{figure}[tp]
  \centering
  \includegraphics[width=0.8\textwidth]{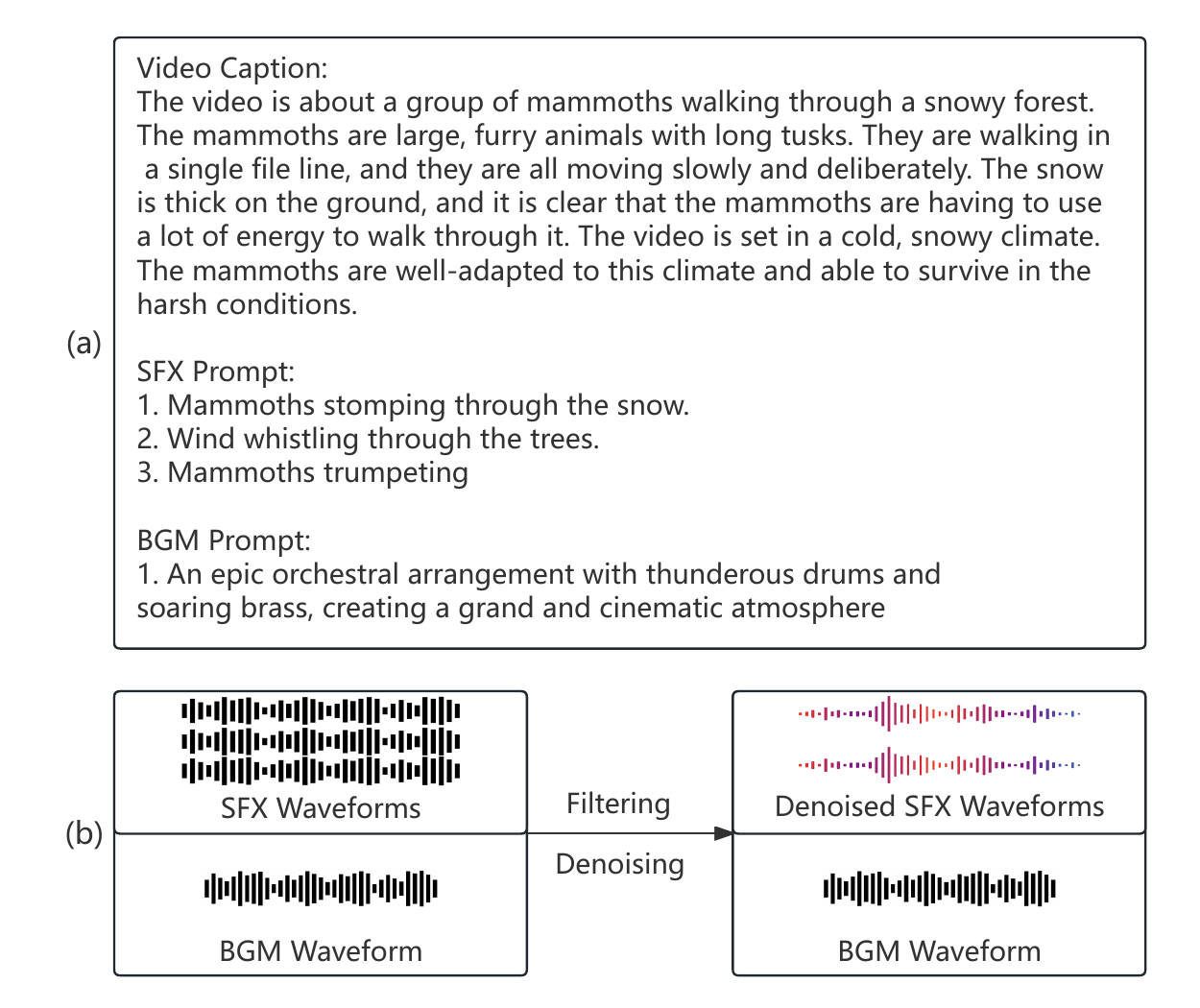}
  \caption{The visualization cases of SVA. Figure \ref{fig:fig4}(a) presents the outputs from MLLM during the prompt generation phase. Figure \ref{fig:fig4}(b) demonstrates the filtering and denoising procedures undertaken during the post-processing phase.}
  \label{fig:fig4}
\end{figure}

In this section, we show a case for generation and post-processing tasks as shown in Figure\ref{fig:fig4}. In Figure \ref{fig:fig4}(a), MLLM provides a video description and formulates an inventive SFX and BGM scheme stored in a dictionary. The dictionary contains two distinct SFX descriptions under the "SFX" key, and one BGM description under the "BGM" key.  These descriptions will serve as prompts, respectively be inputted into AudioGen and MusicGen, generating two SFX and one BGM waveform files. During the post-processing phase as shown in Figure\ref{fig:fig4}(b), one audio track is filtered out and the others are denoised. In the end, the silent video, two SFX tracks, and one BGM track are merged together to create the final complete video. We observed that the majority of the results are of high-quality and semantically consistent, providing the audience with an immersive auditory and visual experience. More samples can be found at \url{https://huiz-a.github.io/audio4video.github.io/}.

\section{Conclusion}
In this work, we propose an efficient and simple framework for audio and music generation based on video semantics. Our framework can achieve efficient video-to-audio generation with natural language as an interface. Instead of training a model from scratch, We utilize easily accessible MLLM to facilitate video-to-text conversion and  open-source audio generation models for text-to-audio conversion, thereby bridging the gap between video and audio. Through our case study, we observed that the video and audio in the final output are semantically aligned, exhibiting high quality and enhancing the overall viewer experience.

Nevertheless, our approach exhibits notable limitations. Utilizing visual semantics as natural language prompts remains a coarse-grained method, which cannot accurately model the intricate relationship between video and audio. As a training-free approach based on MLLM prompt engineering, the evaluation of audio generated by SVA has been temporarily omitted. Our future work will focus on achieving more precise temporal synchronization between video and audio and an exhaustive benchmark.

\section{Discussion}
Video-to-audio generation has been a focal point of research, yielding significant advancements. Diff-Foley\cite{luo2023difffoley} employs a latent diffusion model (LDM) to generate high-quality audio synchronized with video. It leverages contrastive audio-video pre-training to learn temporally aligned features and captures subtle audio-video correlations through cross-attention modules. Seeing-and-Hearing\cite{xing2024seeing} utilizes the pre-trained ImageBind\cite{girdhar2023imagebind} model to align multimodal information in latent space. By calculating losses using the aligner and directly modifying intermediate latent variables, it enhances the correlation between video and audio. Conditional-Foley\cite{du2023conditional} focuses on generating sound effects for repetitive actions like tapping and rubbing. It achieves temporal synchronization and timbre control through self-supervised learning. These works drive progress in multimodal generation and offer new methods for applications such as film editing, virtual reality, and assisting visually impaired individuals. Nonetheless, they are not devoid of limitations.

\textbf{Video-Audio Correlation}. The semantic and temporal correlation between audio and video requires improvement. Some works\cite{xing2024seeing, sheffer2023i} utilize pre-trained models such as ImageBind\cite{girdhar2023imagebind} and Clip\cite{radford2021learning}. These models lack audio-related information in their pre-trained visual features, or learn coarse-grained synchronization, thus limiting the capture of intricate audio-visual correlations. Other studies devise neural networks like attention mechanisms to model the video-audio correlation. However, the performance of these models is constrained by factors such as the quality and quantity of available training data, as well as inherent deficiencies in network design. 

\textbf{Generalization}. Video-to-audio generation e offers extensive applications, facilitating the audio creation for diverse video types. Examples include designing SFX and BGM for short videos, synthesizing speech based on lip movements, and producing Foley for movie scenes. Condition-Foley etc.\cite{du2023conditional, comunità2023syncfusion} aim to produce audio synchronized with repeated actions, which struggle in videos with complex scenes and numerous audio events. The absence of large-scale annotated video-audio datasets is another crucial issue. AudioSet\cite{hershey2021benefit} is a large-scale in-the-wild dataset manually annotated with frame-level annotations of audio events. However, this dataset cannot guarantee the co-occurrence of video and audio, thus directly using raw video-audio pairs may introduce noise. VGGSound\cite{chen2020vggsound} collects numerous video-audio co-occurring clips from YouTube through various audio and image detection methods. Each sample consists of a 10-second video clip with one audio classification label. So the annotation accuracy is relatively low, and it does not support complex scenarios where multiple audio events occur simultaneously. The construction of a dataset that covers diverse scenes, ensures the co-occurrence of video and audio, and maintains fine-grained temporal granularity for audio events is exceptionally challenging.

\textbf{Metrics}. Due to the lack of testing data and evaluation metrics, there is no universal benchmarks for video-to-audio generation. Evaluation metrics employed in recent studies can be broadly categorized into two groups\cite{iashin2021taming, xing2024seeing}. One group assesses the audio fidelity, utilizing metrics such as Inception Score (IS)\cite{Salimans_Goodfellow_Zaremba_Cheung_Radford_Chen_2016} and Frechet Distance (FID)\cite{Heusel_Ramsauer_Unterthiner_Nessler_Hochreiter_2017}. The other group evaluates the video-audio relevance, employing metrics like Mean KL Divergence (MKL)\cite{iashin2021taming} and ImageBind Score (IB)\cite{mei2023foleygen}. At present, there is a scarcity of evaluation metrics for temporal correlation between video and audio. To calculate MKL relies on the audio ground-truth, which is not available when the video inputs are machine-generated. To date, no one has proposed a method to solely rely on visual information for evaluating the temporal correlation. 

Our proposed SVA leverages MLLM to convert the video-to-audio generation task into a text-to-audio generation task. While this approach proves convenient and effective, it lacks the ability to directly harness the rich semantic information in videos and to model the complex video-audio correlation. 

The future direction of video-to-audio generation research should aim to address the aforementioned limitations, including:
\begin{itemize}
    \item Improving the audio fidelity and audio-video semantic and temporal relevance.
    \item Improving model generalization to accommodate various task scenarios.
    \item Developing more effective automatic evaluation metrics. 
    \item Establishing larger-scale datasets and benchmarks.
\end{itemize}

\section{Acknowledgements}
This work was supported in part by the National Key Research and Development Program of China under Grant (2023YFC3310700), the National Natural Science Foundation of China (62202041), the Fundamental Research Funds for the Central Universities (2023JBMC057).

\bibliographystyle{unsrt}  
\bibliography{references}  

\end{document}